\documentstyle[preprint,aps]{revtex}

\begin{document}
\draft
\preprint{KANAZAWA 94-12, July 1994}
\title{
Monopole action and condensation in SU(2) QCD 
}
\author{
Hiroshi Shiba
\footnote{
E mail address: shiba@hep.s.kanazawa-u.ac.jp}
 and Tsuneo Suzuki
\footnote{
E mail address: suzuki@hep.s.kanazawa-u.ac.jp}
}
\address{
Department of Physics, Kanazawa University, Kanazawa 920-11, Japan
}
\date{\today}
\maketitle
\begin{abstract}
An effective monopole action for various extended monopoles 
is derived from vacuum configurations 
after abelian projection in the maximally abelian gauge in $SU(2)$ QCD. 
The action appears to be independent of the lattice volume. 
Moreover it seems to 
depend only on the physical lattice spacing of the renormalized lattice, 
not on $\beta$.
Entropy dominance over energy of monopole loops is seen 
on the renormalized lattice with the spacing 
$b>b_c\simeq 5.2\times10^{-3}
\Lambda_L^{-1}$. This suggests that 
monopole condensation always (for all $\beta$) 
occurs in the infinite-volume limit of 
lattice QCD.
\end{abstract}


\newpage
\narrowtext

To clarify color confinement mechanism is still important in 
particle physics and the effect dual to the Meissner effect is 
believed to be the mechanism \cite{thooft1,mandel}.
This picture is realized in the confinement phase of lattice compact
QED\cite{poly,bank,degrand}.  

To apply the idea to QCD, we have to find a 
color magnetic quantity inside  QCD.  
The 'tHooft idea of abelian
projection of QCD\cite{thooft2} is very interesting.  
The abelian projection of QCD is to extract an abelian gauge theory by 
 performing  a partial gauge-fixing  
such that the maximal abelian subgroup remains unbroken. Then 
QCD can be regarded as an abelian gauge theory
\underline{with magnetic monopoles} and electric charges. 't Hooft 
conjectured that the condensation 
of the abelian monopoles is the confinement mechanism in 
QCD\cite{thooft2}.

There are, however, infinite ways of extracting such an 
abelian theory out of QCD. It seems important to find a 
good gauge in which the conjecture is seen to be realized 
in a coupling region where reliable calculations can be done 
at present.
A gauge called maximally abelian (MA) gauge has been shown to be very 
interesting \cite{kron,yotsu,suzu93}. 
In the MA gauge, 
there are phenomena which may be called abelian dominance
\cite{yotsu,hio}. 
Moreover the monopole currents 
$k_{\mu}(s)$ 
defined similarly as in compact QED\cite{degrand} are 
seen in the MA gauge to be dense and dynamical in the 
confinement phase, while dilute and static in the deconfinement phase
\cite{suzu93}.
Especially the string tension can be reproduced by the monopole 
contributions alone\cite{shiba4,stack2}.

Although these data support the 'tHooft conjecture, there is not a direct 
evidence of abelian monopole condensation in the QCD vacuum.
In the case of compact QED, the exact dual transformation can be done and 
leads us to an action describing a monopole Coulomb gas,
when one adopt the partition function of the Villain form
\cite{bank,peskin,frolich,smit}. 
Monopole condensation is shown to 
occur in the confinement phase from energy-entropy balance
of monopole loops. 

In the case of QCD, however, we encounter a difficulty 
in performing the exact dual transformation. Hence we tried to do 
it numerically first by fixing an effective U(1) action. Namely, 
the abelian dominance suggests that a set 
of $U(1)$ invariant operators 
are enough to describe confinement after the abelian projection. 
Then there must exist an effective $U(1)$ action 
describing confinement. As reported in \cite{suzu93}, however,
to derive a compact form of such a $U(1)$ action 
in terms of abelian link variables was unsuccessful in the scaling region.
Here we propose a method to directly 
determine a monopole action numerically from the vacuum configurations 
of monopole currents 
 given by 
Monte-Carlo simulations in the MA gauge\cite{shiba2,shiba3}. 
This can be done by extending the Swendsen method\cite{swendsen}
 which was developed in the 
studies of the Monte-Carlo renormalization group. We have shown recently that 
the method is successfully applied to compact QED\cite{shiba1}.
When applied to SU(2) QCD in the MA gauge, 
it is suggested that 
entropy dominance over energy and then monopole condensation always occur 
in the infinite-volume limit of lattice QCD. 
A preliminary report was published 
in \cite{shiba2,shiba3}.

The MA gauge is given on a lattice  by performing
 a local gauge transformation 
\begin{equation}
\widetilde{U}(s,\mu)=V(s)U(s,\mu)V^{-1}(s+\hat\mu)
\end{equation}
such that a quantity
\begin{equation}
  R=\sum_{s,\hat\mu}{\rm Tr}\Big(\sigma_3 \widetilde{U}(s,\mu)
        \sigma_3 \widetilde{U}^{\dagger}(s,\mu)\Big) \label{R}
\end{equation}
is maximized\cite{kron}. 
In the continuum limit, this corresponds to a gauge 
$(\partial_{\mu}+igA_{\mu}^3)(A_{\mu}^1-iA_{\mu}^2)=0.$
Such a gauge fixing keeps a residual U(1) invariance.

 After  the  gauge fixing is done, an abelian link gauge field 
 is extracted from the $SU(2)$ link variables as follows;
\def\absc{(1-\vert c(s,\mu)\vert^2)^{1/2}}
\begin{eqnarray}
 \widetilde{U}(s,\mu) = \left( \begin{array}{cc}
         \absc        & -c^*(s,\mu) \\
                  c(s,\mu) &  \absc           
\end{array} \right)
\left( \begin{array}{cc}
u(s,\mu) & 0 \\
0 & u^*(s,\mu) 
\end{array} \right),
\end{eqnarray}
where $u(s,\mu)= \exp(i\theta(s,\mu))$ represents the abelian link field 
and $c(s,\mu)$ corresponds to charged matter fields.

A plaquette variable $\theta_{\mu\nu}(s)\in(-4\pi,4\pi)$ is given 
by a link angle $\theta(s,\mu)\in[-\pi,\pi)$ 
as $\theta_{\mu\nu}(s)=\theta(s,\mu)+
\theta(s+\hat\mu,\nu)-\theta(s+\hat\nu,\mu)-\theta(s,\nu)$.
The plaquette variable 
$\theta_{\mu\nu}(s)$ can be decomposed into 
\begin{equation}
\theta_{\mu\nu}(s)= \bar{\theta}_{\mu\nu}(s) +2\pi m_{\mu\nu}(s),
\end{equation}
where $\bar{\theta}_{\mu\nu}(s)\in [-\pi,\pi)$ is 
interpreted as the electromagnetic flux through the plaquette 
and $m_{\mu\nu}(s)$ 
can be regarded as a number of the Dirac string penetrating the plaquette.
DeGrand and Toussaint\cite{degrand} defined a monopole current as 
\begin{eqnarray}
k_{\mu}(s) & = & \frac{1}{2}\epsilon_{\mu\nu\rho\sigma}
\partial_{\nu}m_{\rho\sigma}(s+\hat{\mu}), \label{dgcur} 
\end{eqnarray}
where $\partial$ is the forward difference. 
It is defined on a link of the dual lattice 
(the lattice with origin shifted by 
half a lattice distance all in four directions).
$k_{\mu}(s)$  can take only the values $0, \pm 1, \pm 2$ and satisfies 
\begin{equation}
\partial'_{\mu}k_{\mu}(s)=0,
\end{equation}
where $\partial'$ is the backward derivative on the dual lattice.

\begin{figure}
\vspace{2cm}
\epsfxsize=0.6\textwidth
\begin{center}
\leavevmode
\epsfbox{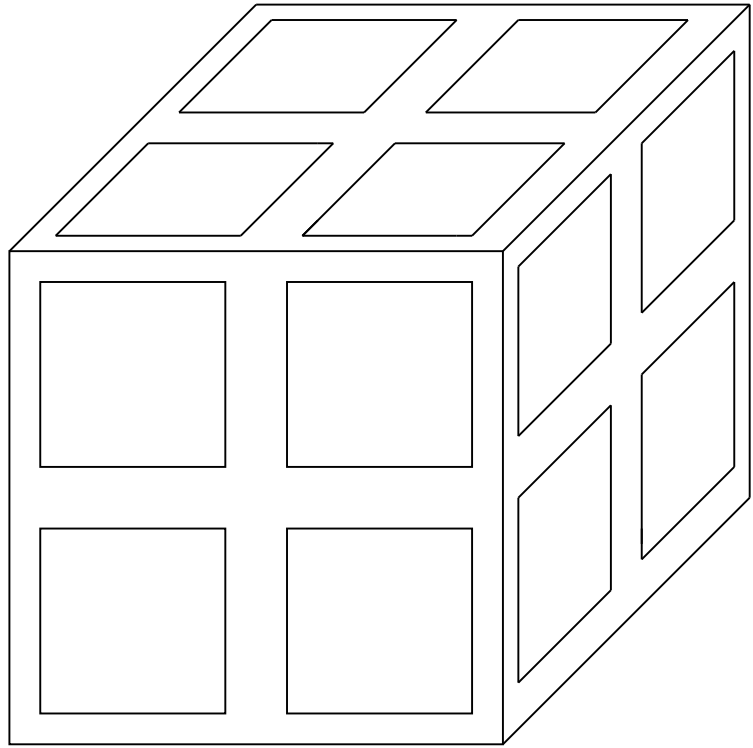}
\end{center}
\vspace{2cm}
\caption{
An extended cube on which an $2^3$ extended monopole is defined as the sum of 
eight (=$2^3$) smallest monopoles. 
}
\label{fig1}
\end{figure}

Long-distance behaviors are expected to be important in the 
confinement phase of QCD. 
Hence we consider extended monopoles of the type-2
\cite{ivanenko} as shown in Fig.\ \ref{fig1}.
\footnote{
Type-1 extended 
monopoles\cite{ivanenko} are not adopted 
here, since the separation of the Dirac string 
is problematic. 
}
The $n^3$ extended monopole of the type-2 has a total magnetic 
charge inside the $n^3$ cube and is defined on 
a sublattice with the spacing
$b=na$,  $a$ being the spacing of the original lattice.
\begin{eqnarray}
k_{\mu}^{(n)}(s) &=& \frac{1}{2}\epsilon_{\mu\nu\rho\sigma}
\partial_{\nu}m_{\rho\sigma}^{(n)}(s+\hat{\mu}) \label{excur}\\ 
    & = & \sum_{i,j,l=0}^{n-1}k_{\mu}(ns+(n-1)\hat{\mu}+i\hat{\nu}
     +j\hat{\rho}+l\hat{\sigma}),\\
m_{\rho\sigma}^{(n)}(s) &=& \sum_{i,j=0}^{n-1} 
m_{\rho\sigma}(ns+i\hat{\rho}+j\hat{\sigma}).
\end{eqnarray}
The definition of the type-2 extended monopoles 
corresponds to making a block
spin transformation of the monopole currents with the scale factor $n$.
We call the sublattice as a renormalized lattice.
We derived the coupling constants for $1^3, 2^3, 3^3$, and $4^3$ 
extended monopoles.

A theory of monopole loops is given in general by 
the following partition 
function  
\begin{equation}
Z= (\prod_{s,\mu}\sum_{k_{\mu}(s)=-\infty}^{\infty})
(\prod_s \delta_{\partial'_{\mu}k_{\mu}(s),0})
\exp(-S[k]), \label{zmono2}
\end{equation}
where $k_{\mu}(s)$ 
is the conserved integer-valued monopole current defined above in
(\ref{dgcur}) or (\ref{excur}) and $S[k]$ is a monopole action describing 
the theory. 
Consider a set of all independent operators which are summed up 
over the whole lattice. We denote each operator as $S_i [k]$. Then 
the action can be written as a linear combination of these operators:
\begin{equation}
S[k] = \sum_i f_i S_i [k], 
\end{equation}
where $f_i$ are coupling constants.

Let us now determine the monopole action using the monopole 
current ensemble which are calculated  from vacuum 
configurations generated by Monte-Carlo simulations.
Since the dynamical variables here are $k_{\mu}(s)$ satisfying the 
conservation rule, it is necessary to extend the 
original Swendsen method\cite{swendsen}. 
Consider a plaquette $(s',\hat{\mu'},\hat{\nu'})$ instead of a link 
on the dual lattice.
Introducing a new set of coupling constants 
$ \{\tilde{f}_i\} $, define
\begin{equation}
\bar{S}_i [k]= \frac{\sum_{M=-\infty}^{\infty}S_i[k']
\exp(-\sum_i \tilde{f}_i S_i [k'])}
{\sum_{M=-\infty}^{\infty}\exp(-\sum_i \tilde{f}_i S_i [k'])},
\label{sbar}
\end{equation}
where 
$k'_{\mu}(s)  =  k_{\mu}(s) + M(\delta_{s,s'}\delta_{\mu\mu'}
+ \delta_{s,s'+\hat{\mu}'}\delta_{\mu\nu'}  
- \delta_{s,s'+\hat{\nu}'}\delta_{\mu\mu'}
- \delta_{s,s'}\delta_{\mu\nu'}).$
When all $\tilde{f}_i$ are equal to $ f_i $,
one can prove an equality   
$ \langle \bar{S}_i \rangle  =  \langle S_i \rangle $,
where the expectation values are taken 
over the above original action 
with the coupling constants $\{f_i\}$:
\begin{eqnarray}
\langle O[k]\rangle =
\frac{
(\prod_{s,\mu}\sum_{k_{\mu}(s)=-\infty}^{\infty})
(\prod_s \delta_{\partial'_{\mu}k_{\mu}(s),0}) O[k]
\exp(-\sum_{i}f_i S_i [k])}
{(\prod_{s,\mu}\sum_{k_{\mu}(s)=-\infty}^{\infty})
(\prod_s \delta_{\partial'_{\mu}k_{\mu}(s),0})
\exp(-\sum_{i}f_i S_i [k])}. \label{ok1}
\end{eqnarray}
When there are some $\tilde{f}_i$ not equal to $ f_i $,
 one may expand the difference as follows  
\begin{equation}
\langle \bar{S}_i - S_i \rangle = 
\sum_j 
\langle \overline{S_i S_j}-\bar{S}_i\bar{S}_j \rangle
(f_j - \tilde{f}_j)    \label{ssbar}, 
\end{equation}
where only the first order terms are written down.
This allows an iteration scheme for determination 
of the unknown constants
$f_i$. For more details, see \cite{shiba1} where 
(\ref{ssbar}) is derived explicitly 
and is applied to compact QED.

Practically we have to restrict the number of 
interaction terms.
\footnote{
All possible types of interactions are not independent, since 
$\partial'_{\mu}k_{\mu}(s)=0$. We can get rid of almost 
all interactions between 
different components of the currents from the quadratic action 
by use of the conservation rule.
}
We  adopted
12 types of quadratic interactions listed in Table \ref{table1} 
in most of these 
studies.
To check the validity of the truncation, 
we  also included more quadratic
and even quartic interactions in some cases, but 
found little difference.

Monte-Carlo simulations 
were done on $8^4, 12^4, 14^4, 16^4, 
18^4, 20^4$, and $24^4$ lattices for 
$0 < \beta \le 3.0$. 

\begin{figure}[tb]
\epsfxsize=\textwidth
\begin{center}
\leavevmode
\epsfbox{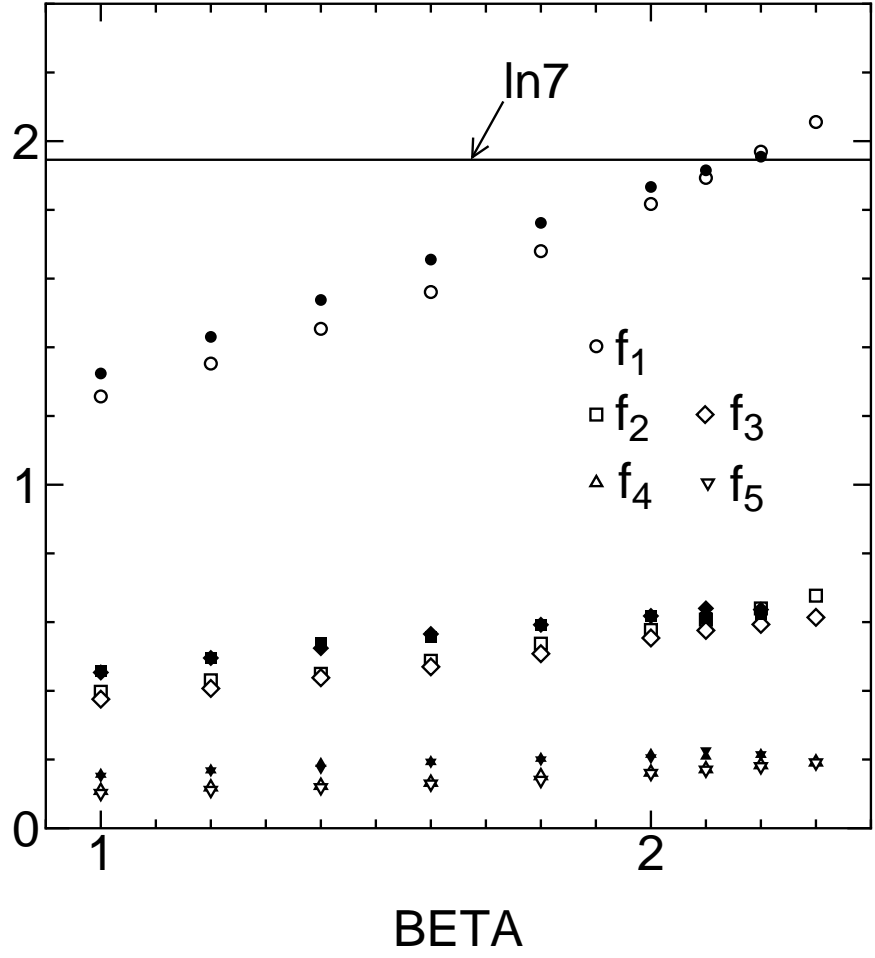}
\end{center}
\caption{Coupling constants $f_i$ versus $\beta$ in the strong coupling 
region. White symbols denote the $SU(2)$ data, whereas black 
ones show the data of the Wilson form of compact QED with the substitution 
$\beta_w = \beta/2$.
}
\label{fig2}
\end{figure}

\begin{figure}[tb]
\epsfxsize=\textwidth
\begin{center}
\leavevmode
\epsfbox{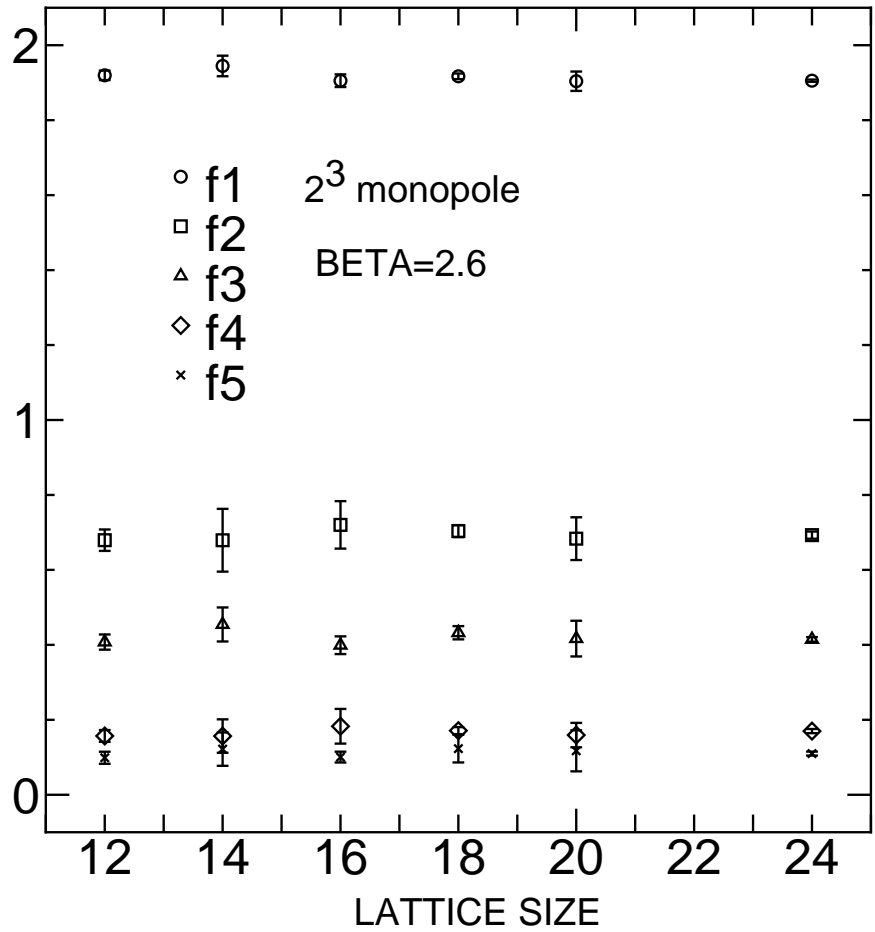}
\end{center}
\caption{Coupling constants $f_i$ versus lattice size . 
}
\label{fig3}
\end{figure}

\begin{figure}[tb]
\epsfxsize=\textwidth
\begin{center}
\leavevmode
\epsfbox{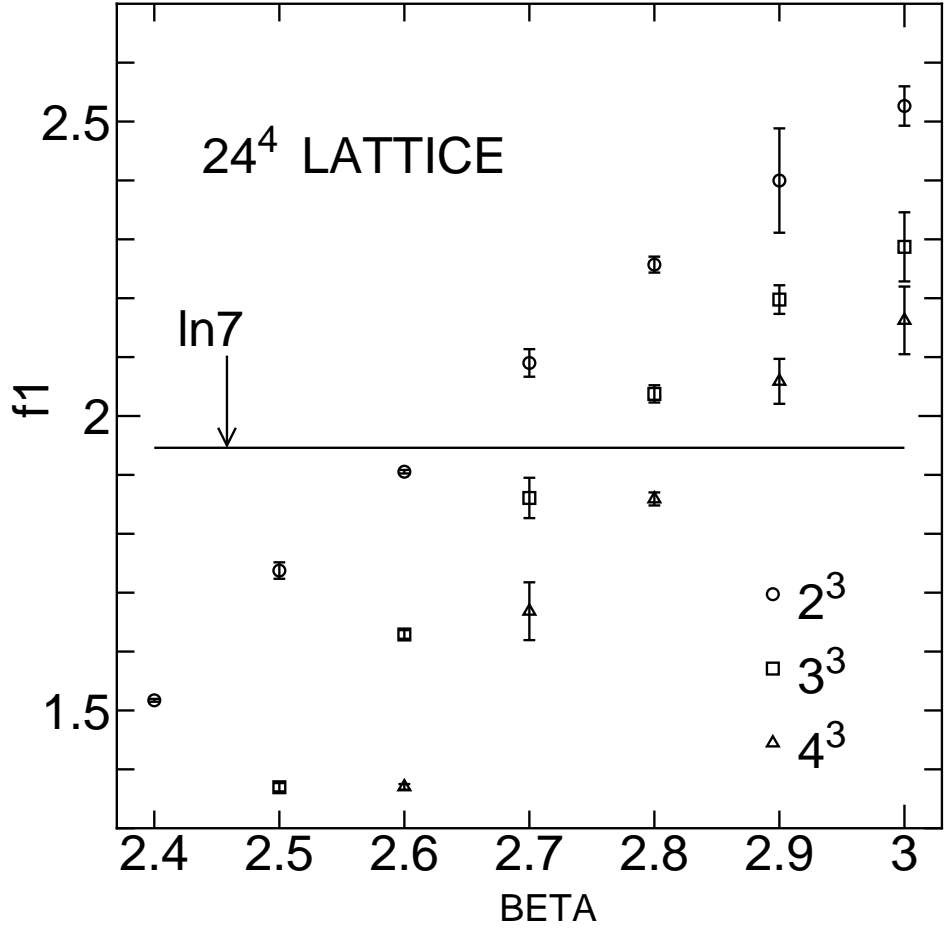}
\end{center}
\caption{Coupling constants $f_1$ versus $\beta$ for 
$2^3,3^3$, and $4^3$ 
extended monopoles on $24^4$ lattice. 
}
\label{fig4}
\end{figure}

Our results are summarized in the following.
\begin{enumerate}

\item
The coupling constants $f_1 \sim f_5$ for $1^3$ monopoles are plotted 
(white symbol)
for small $\beta$ in Fig.\ \ref{fig2}. As shown in
 \cite{suzu93}, an effective $U(1)$ action is determined for small $\beta$.
It takes a Wilson $U(1)$ action with the effective coupling constant 
$\beta_w = \beta/2 $.  In Fig.\ \ref{fig2}, we have also plotted the data 
(black symbol)
obtained in the case of the Wilson $U(1)$ action\cite{shiba1} 
with the coupling constant substitution.
Equivalence of the two theories for small $\beta$ 
is seen also in the form of the monopole 
action.

\item
For $\beta \ge 2.4$, extended monopoles are considered.
\footnote{
In the case of compact QED, a monopole action could be determined only for 
$1^3$ monopoles. For extended monopoles, the iteration to fix the action 
was not convergent.
}
The coupling constants $f_i$ are fixed for 
not so large value of
$b$ ,i.e., $b < 1.3\times 10^{-2}(\Lambda_L)^{-1}$. 
The first important observation is that 
the coupling constants determined are almost 
independent of the lattice
volume as seen from Fig.\ \ref{fig3}. 
We see $f_1$ is dominant and the coupling constants
decrease rapidly as the distance between 
the two monopole currents increases.

\item
Since the monopole action is determined, let us consider energy-entropy 
balance of monopole loops as done in compact QED\cite{bank}.
Monopole world lines form closed loops on the dual lattice.
Define $n_L [k]$ as the number of monopole loops of length $L$
in a monopole configuration $\{k_{\mu}(s)\}$. The contribution of 
monopole loops of length $L$ is represented by the average of 
$n_L [k]$ :
\begin{equation}
\langle n_L [k] \rangle =\frac{\sum_{\{k\}}n_L [k]e^{-S[k]}}
{\sum_{\{k\}}e^{-S[k]}}, \label{nk}
\end{equation}
where 
\begin{equation}
\sum_{\{k\}}=
(\prod_{s,\mu}\sum_{k_{\mu}(s)=-\infty}^{\infty})
(\prod_s \delta_{\partial'_{\mu}k_{\mu}(s),0}). 
\end{equation}
Monopole loops of length $L$ take various forms. Choose a 
monopole loop $C_L$ of length $L$ and define 
\begin{equation}
\delta_{C_L}[k]= \left\{ \begin{array}{ll}
1 & \mbox{if $C_L \in \{k\}$}\\
0 & \mbox{otherwise\ . }
\end{array} \right.
\end{equation}
$n_L[k]$ is written as 
\begin{equation}
n_L[k]=\sum_{C_L}\delta_{C_L}[k],
\end{equation}
where $\sum_{C_L}$ denotes the sum over all loops of length $L$.
Then the average (\ref{nk}) is 
\begin{equation}
\langle n_L [k] \rangle = \sum_{C_L}\langle \delta_{C_L}[k] \rangle.
\label{avern1}
\end{equation}
$\langle \delta_{C_L}[k] \rangle$ depends not only on the loop length $L$ 
but also on the shape of $C_L$. 

One may define the 'energy' $E(L)$ of monopole loops of length $L$ as 
the average of $\langle \delta_{C_L}[k] \rangle$ over all 
loops of length $L$:
\begin{eqnarray}
e^{-E(L)} &=& \frac{\sum_{C_L}\langle \delta_{C_L}[k] \rangle}
{\sum_{C_L} 1}, \\ 
\langle \delta_{C_L}[k] \rangle &=& \frac{\sum_{\{k\}}\delta_{C_L}[k]
e^{-S[k]}}{\sum_{\{k\}}e^{-S[k]}}. 
\end{eqnarray}
Then we get  
\begin{equation}
\langle n_L [k] \rangle = \rho (L)e^{-E(L)},
\label{avern2}
\end{equation}
where $\rho (L)= \sum_{C_L} 1$ is the total number of monopole loops 
of length $L$ on a dual lattice and ln$\rho (L)$ may be regarded as the 
'entropy' of monopole loops of length $L$.

Now we write 
\begin{eqnarray}
\langle \delta_{C_L}[k] \rangle = 
e^{-S(C_L)} \times \frac{\sum_{\{k\}}\delta_{C_L}[k]
e^{-S[k]+S(C_L)}}{\sum_{\{k\}}e^{-S[k]}}, 
\end{eqnarray}
where $S(C_L)$ is the contribution of a loop $C_L$ alone to the monopole 
action. For large $L$, monopole currents in a loop are distributed 
randomly in average. If the action consists of quadratic terms alone as 
assumed in our analysis, interaction terms between different currents would 
cancel. $S(C_L)$ may be approximated by the self-energy part:
\begin{equation}
S(C_L) \sim f_1 \times L,
\end{equation}
where we assumed the dominance of currents with a unit charge 
$|k_{\mu}(s)|=1$. 
\footnote{
We have made a histogram analysis of the 
magnetic charge distribution 
of each extended monopole. The case 
with $|k_{\mu}(s)|= 2$ is much 
(less than 5\%) suppressed in comparison 
with the $|k_{\mu}(s)|=1$ case 
and charges with $|k_{\mu}(s)|\geq 3$ barely appear even 
for $4^3$ extended monopoles in the confinement 
phase. 
}

On the otherhand, $\sum_{\{k\}}\delta_{C_L}[k]e^{-S[k]+S(C_L)}$ has two types 
of contributions, i.e., 1)those from other loops in the presence of $C_L$ and 
2)the interactions between $C_L$ and other loops. 
The latter may be cancelled again from randomness when $L$ is large.
When we fix $C_L$ and make the lattice volume larger, the ratio of 
the number of links occupied by $C_L$ to total link number becomes smaller.
Hence the $C_L$ dependence of the former may be negligible 
in the infinite-volume limit.  We may approximate for large $L$
\begin{eqnarray}
\langle \delta_{C_L}[k] \rangle \propto 
e^{-f_1 L},
\end{eqnarray}
which leads to 
\begin{equation}
E(L) \sim f_1 L + const. \label{elf}
\end{equation}
  
If we consider non-backtracking random walks on a four 
dimensional dual lattice,
the number of loops of length $L$ through a given site behaves asymptotically 
as $7^L$ for large $L$. We get 
\begin{equation}
\rho (L) \propto 7^L. \label{rho}
\end{equation}
From (\ref{elf}) and (\ref{rho}), we obtain 
\begin{eqnarray}
\langle n_L [k] \rangle \propto 
e^{(\ln 7 -f_1)L} \label{nl3}
\end{eqnarray}
for large $L$.
If $f_1 <$ ln7, the 'entropy' dominates over the 'energy' 
and the long monopole loops become dominant. Namely the 
monopole condensation occurs. The same thing happens also in compact 
QED\cite{bank}.

We plot $f_1$ versus $\beta$ for various 
extended monopoles  on $24^4$ lattice
in comparison with the entropy value ln7 
for the infinite volume in Fig.\ \ref{fig4}.
Each extended monopole has its own $\beta$ region where 
the condition  $f_1 <$ ln7 is satisfied. When  the extendedness 
is bigger, larger $\beta$ is included in such a region. 

\item
The behaviors of the coupling constants are 
different for different extended 
monopoles. But if we plot them versus $b$ , 
we get a unique curve as in Figs.\ \ref{fig5}, \ref{fig6} and \ref{fig7}. The 
coupling constants seem to depend only on $b$, not on the extendedness 
nor $\beta$.  This suggests the existence 
of the continuum limit and  the monopole action 
in the continuum may be similar to that given here. 
From Fig.\ \ref{fig5}, 
we see the self-energy per monopole loop length $f_1$ 
on the renormalized lattice decreases as 
the length scale $b$ increases.
A critical length $b_c \sim 5.2\times 10^{-3}(\Lambda_L)^{-1}$ exists 
at which the $f_1$ value crosses the ln7 line.

Together with  
the first observation of the volume independence,  
we may get the following important conclusion in 
 the infinite-volume limit. 
One can take always (for all $\beta$) a renormalized 
lattice with a spacing $b > b_c$ by making a block spin 
transformation of the monopole currents with a sufficiently
large scale factor. Hence 
\underline {the QCD vacuum in the infinite-volume limit of lattice QCD 
is always (for all $\beta$)}
\underline { in the monopole condensed phase.} 
This is the most 
important observation of this report.

\item
What is the meaning of the critical length $b_c$? 
When the lattice distance $b$ of the renormalized lattice is larger than 
the critical length $b_c$, the condensation of monopole loops is 
shown from the above 'energy-entropy' balance. 
If monopoles are physical quantities, it must have a finite energy and a 
finite physical size $R_m$.
\footnote{
Note that $b$ represents 
a scale at which behaviors of the 
monopole currents are examined. One should 
not confuse it with the physical 
monopole size $R_m$.
}
 When the lattice distance $b$  is less than the 
physical monopole size $R_m$, one can not regard monopoles as a point particle
running a world line on such a renormalized lattice. 
Hence the 'energy' and the 'entropy' of such monopoles 
can not be represented as in (\ref{elf}) and (\ref{rho}). 
The condition  $f_1 <$ ln7
can not be applicable to such monopoles. 
The above consideration suggests $R_m \sim O(b_c)$. 

\item
Using the data of $f_i$, we try to fix the explicit form of the monopole
action. Although the coupling constants for largely separated currents 
are too small with large errors, the monopole action may be fitted by
 the action  predicted 
theoretically by Smit and Sijs\cite{smit}:
\begin{eqnarray}
S[k] & = & \sum m_0 b k_{\mu}(s)k_{\mu}(s) 
+ \frac{1}{2}(\frac{4\pi}{g(b)})^2 \sum 
k_{\mu}(s)\bar{D}(s-s')k_{\mu}(s'), \label{action}
\end{eqnarray}
where $g(b)$ is the SU(2) running coupling constant
\begin{eqnarray}
g(b)^{-2}= \frac{11}{24\pi^2}\ln(\frac{1}{b^2\Lambda^2}) 
+ \frac{17}{44\pi^2}
\ln\ln(\frac{1}{b^2\Lambda^2}).
\end{eqnarray}
The scale parameter determined here is 
$\Lambda \sim 42\Lambda_L$ which is different from that in \cite{smit}.
$\bar{D}(s)$ is a 
modified lattice Coulomb propagator\cite{smit}.
It is almost equal to the usual lattice one except 
at zero and one lattice distances where the finite 
monopole size gives a modification.
The first can be absorbed into a redefinition of the 'mass' term $m_0$. 
The existence of the finite monopole size leads to two different $\bar{D}(1)$ 
when the two currents are separated on the same line or on the parallel line. 
This may explain the data $f_2 \neq f_3$\cite{smit}. 
The solid lines in Fig.\ \ref{fig5}, Fig.\ \ref{fig6} and Fig.\ \ref{fig7} 
are the prediction given 
by the action with the parameters 
written in the figures.
The existence of the 'monopole mass' 
$m_0$ and the running coupling 
constant $g(b)$ is characterisitic of the  
action of SU(2) QCD in comparison with that of 
compact QED.

\item
The action (\ref{action}), if actually correct, depends only on $b$, not 
on the extendedness of the monopole currents. As shown above, 
considering the extended monopoles 
corresponds to making a block spin transformation on a dual lattice.
Hence the action (\ref{action}) is independent of the block spin 
transformation. This is similar conceptually 
to  the perfect lattice 
action by Hasenfratz and Niedermayer\cite{hasen}. 

\item
It is interesting to see the relation between 
the monopole dynamics and 
the deconfinement transition in a finite volume\cite{kovac}. 
This will be discussed in a separate paper\cite{kita}.
\end{enumerate}

In summary, the monopole condensation is 
shown, in the maximally abelian gauge, to occur always 
in the infinite-volume limit. The abelian charge is confined due to the dual 
Meissner effect. The U(1) charge confinement after abelian projection 
means color confinement in SU(2) QCD as proved in \cite{thooft2}.

However we have not here clarified 
the origin of $\beta_c$ due to the finite-size effect 
in terms of abelian monopoles. This will be published elsewhere\cite{kita}. 
In addition, 
we have not yet understood what happens in other abelian projections. 
If abelian monopoles play an essential role in the continuum, 
the mechanism of monopole condensation may not depend on a gauge choice.
This will be studied in future.

We wish to acknowledge Yoshimi Matsubara for 
useful discussions especially on
the monopole dynamics and the finite-size effects. Also we are thankful to  
 Osamu Miyamura and Shinji 
Hioki for discussions on the perfect action on a lattice.
This work is financially supported by JSPS Grant-in Aid for 
Scientific  Research (B)(No.06452028).

\begin{table}
\caption{
The quadratic terms of the monopole action adopted. 
All terms in which 
the relation of the two currents is equivalent are added 
to make each $S_i [k]$ invariant under translation and rotation. 
Here 
$\hat{a} \equiv \hat{\nu}+\hat{\rho}$, 
$\hat{b} \equiv \hat{\nu}+\hat{\rho}+\hat{\omega}$,
$\hat{c} \equiv \hat{\mu}+\hat{\nu}+\hat{\rho}+\hat{\omega}$,
$\hat{d} \equiv 2\hat{\mu}+\hat{\nu}$,
where $\hat{\mu}, \hat{\nu}, \hat{\rho}$ and $\hat{\omega}$
denote unit vectors in four different directions.
\label{table1}}
\begin{tabular}{rcr}
 i   &   $S_i [k]$   &    \\
\tableline
 1&$k_{\mu}(s)k_{\mu}(s)                    $& \\
 2&$k_{\mu}(s)k_{\mu}(s+\hat{\mu})          $& \\
 3&$k_{\mu}(s)k_{\mu}(s+\hat{\nu})          $& \\
 4&$k_{\mu}(s)k_{\mu}(s+\hat{\mu}+\hat{\nu})$& \\
 5&$k_{\mu}(s)k_{\mu}(s+\hat{a})            $& \\
 6&$k_{\mu}(s)k_{\mu}(s+2\hat{\mu})         $& \\
 7&$k_{\mu}(s)k_{\mu}(s+\hat{\mu}+\hat{a})  $& \\
 8&$k_{\mu}(s)k_{\mu}(s+\hat{b})            $& \\
 9&$k_{\mu}(s)k_{\mu}(s+\hat{c})            $& \\
10&$k_{\mu}(s)k_{\mu}(s+\hat{d})            $& \\
11&$k_{\mu}(s)k_{\nu}(s+\hat{d})            $& \\
12&$k_{\mu}(s)k_{\mu}(s+\hat{d}+\hat{\rho}) $& \\
\end{tabular}
\end{table}
   
\input epsf   
   
\begin{figure}[tb]
\epsfxsize=\textwidth
\begin{center}
\leavevmode
\epsfbox{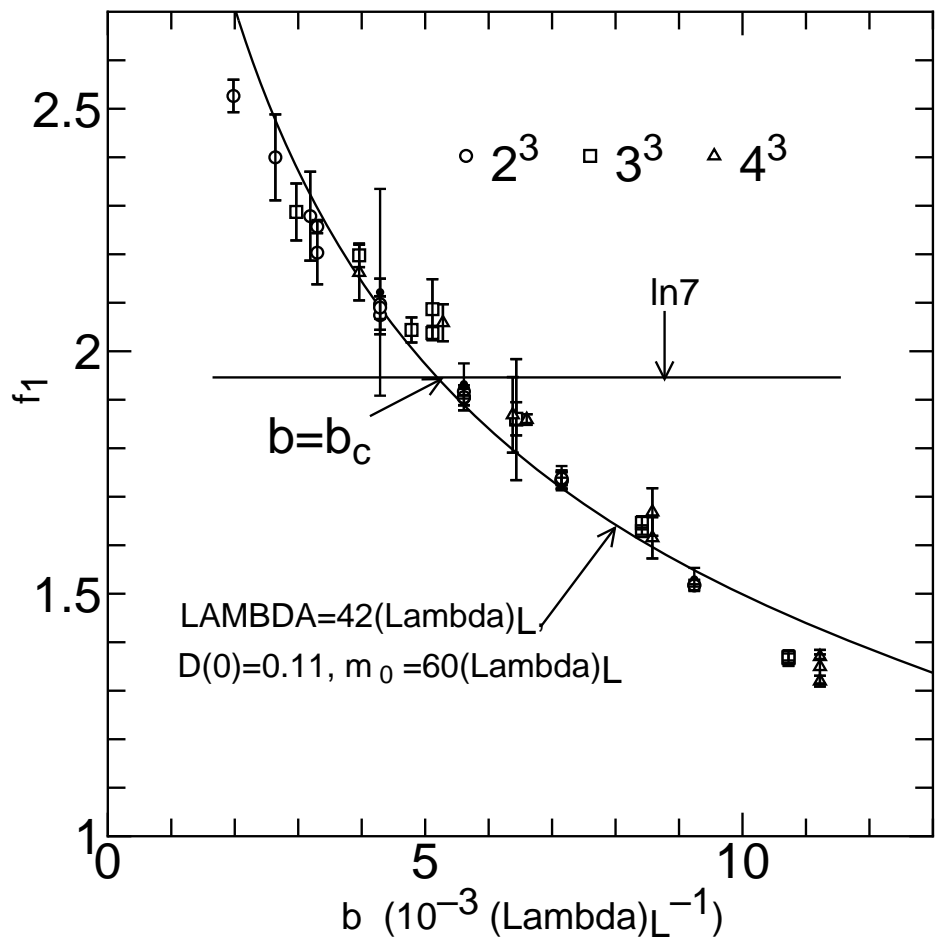}
\end{center}
\caption{Coupling constants $f_1$ versus $b$. }
\label{fig5}
\end{figure}

\begin{figure}[tb]
\epsfxsize=\textwidth
\begin{center}
\leavevmode
\epsfbox{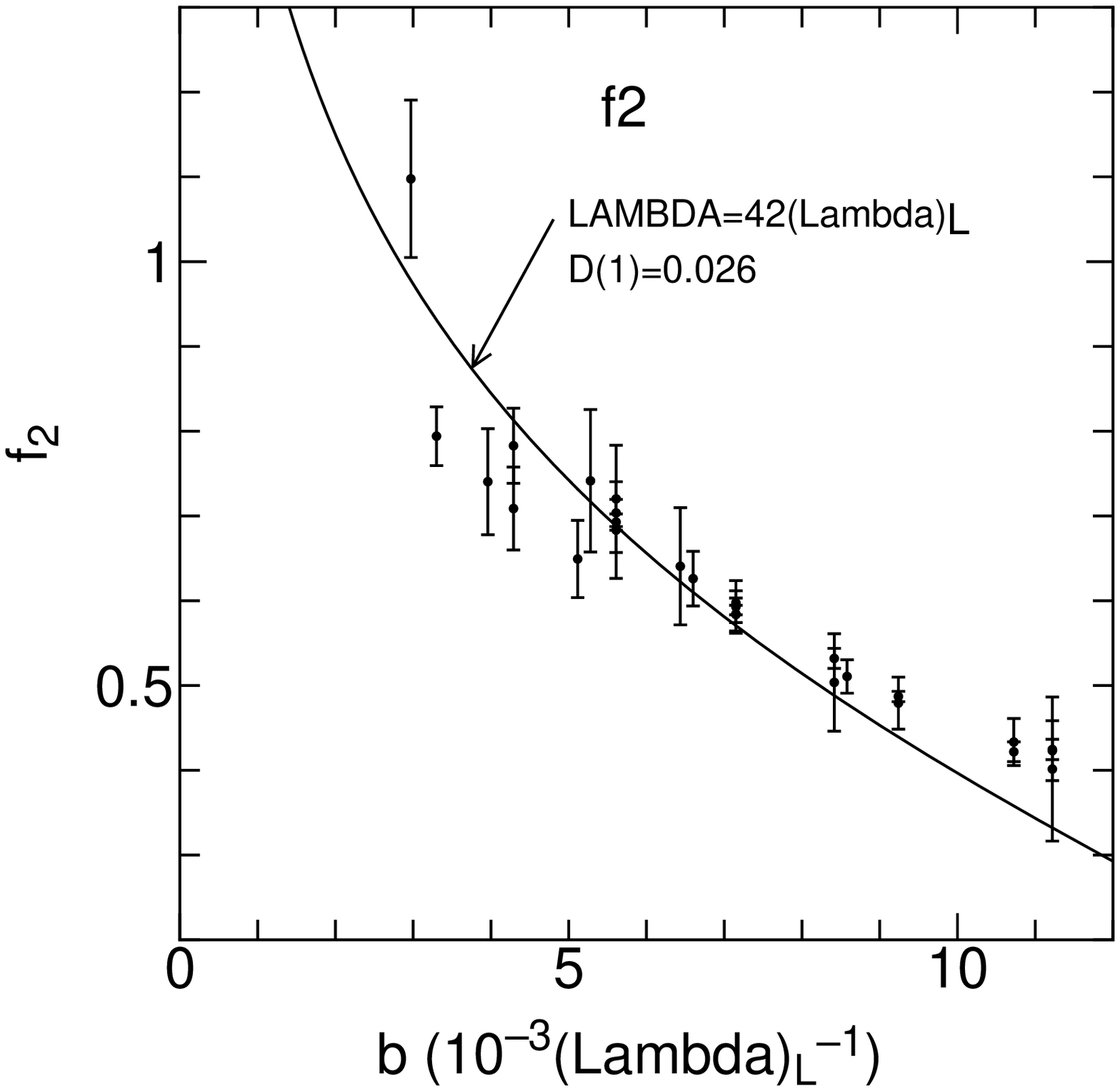}
\end{center}
\caption{Coupling constants $f_2$ versus $b$. 
The data with too large errors are omitted.
}
\label{fig6}
\end{figure}

\begin{figure}[tb]
\epsfxsize=\textwidth
\begin{center}
\leavevmode
\epsfbox{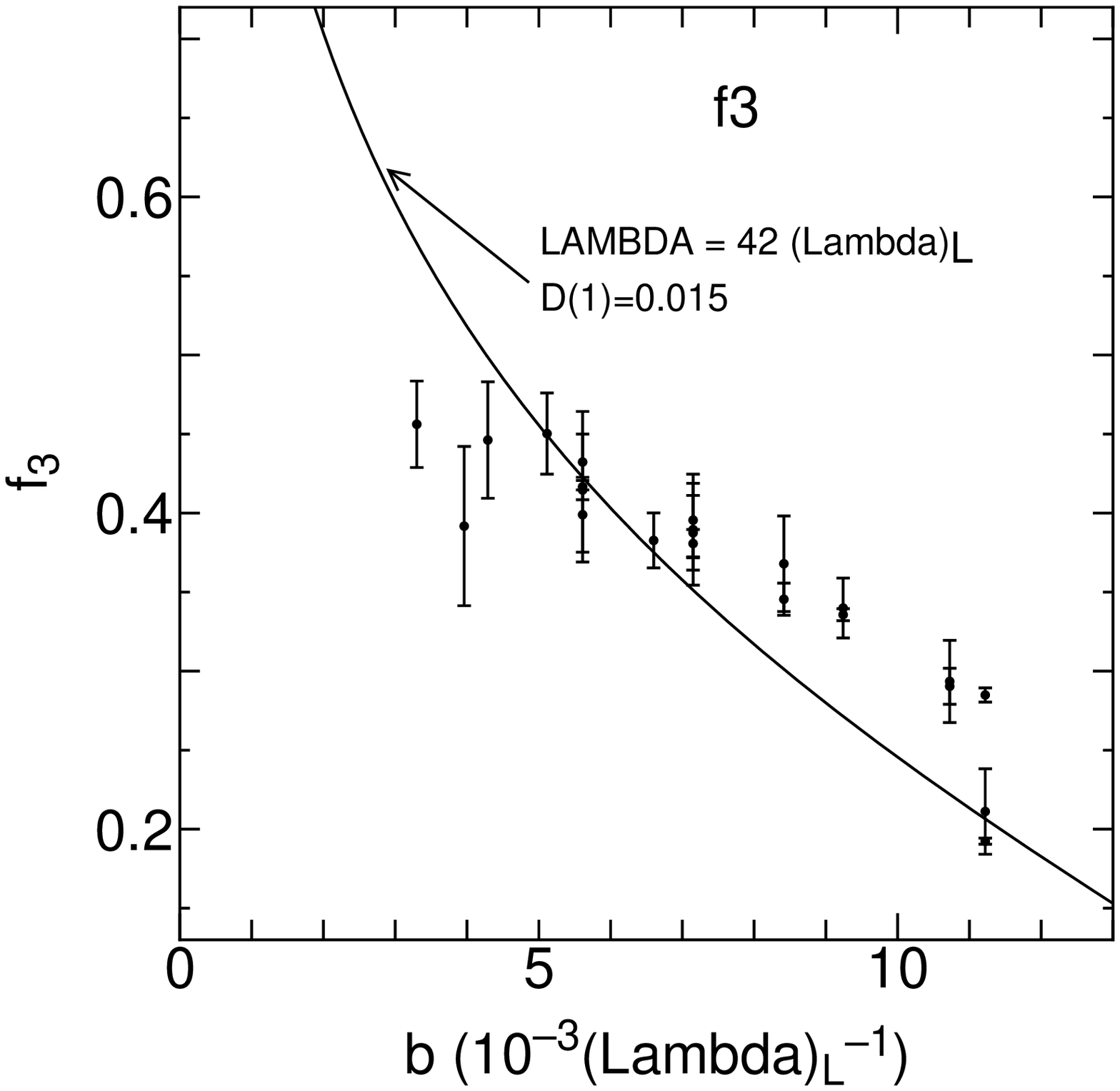}
\end{center}
\caption{Coupling constants $f_3$ versus $b$. 
The data with too large errors are omitted.
}
\label{fig7}
\end{figure}

\end{document}